\documentclass{article}

\usepackage{arxiv}

\usepackage[utf8]{inputenc} 
\usepackage[T1]{fontenc}    
\usepackage{hyperref}       
\usepackage{url}            
\usepackage{booktabs}       
\usepackage{amsfonts}       
\usepackage{nicefrac}       
\usepackage{microtype}      
\usepackage{lipsum}
\usepackage{graphicx}
\usepackage{tcolorbox}
\usepackage{xcolor}
\graphicspath{ {./images/} }

\newtcolorbox{hypothesis}[1]{
    colbacktitle=gray!20,  
    coltitle=black,        
    colback=gray!10,       
    colframe=gray!20,        
    fonttitle=\bfseries,
    title={#1},
     before skip=15pt, 
    after skip=15pt 
}

\title{BDD-Based Framework with RL Integration: An approach for videogames automated testing}

\author{
 Vincent Mastain \\
  Graduate Student, CY Tech\\
  Cergy, France \\
  \texttt{vincentmastain@gmail.com} \\
   \And
   Fabio Petrillo \\
 Departement de génie logiciel \\
 Ecole de Technologie Supérieure, ÉTS  \\ 
 Montreal, QC \\
  \texttt{fabio.petrillo@etsmtl.ca} \\
}

\begin{document}
\maketitle
\begin{abstract}
Testing plays a vital role in software development, but in the realm of video games, the process differs from traditional software development practices. Game developers typically rely on human testers who are provided with checklists to evaluate various elements. While major game developers already employ automated testing using script-based bots, the increasing complexity of video games is pushing the limits of scripted solutions, necessitating the adoption of more advanced testing strategies. To assist game studios in enhancing the quality of their games through  automated testing, we propose the integration of Behavior Driven Development (BDD) with Reinforcement Learning (RL). 
This positional paper summarizes our proposal and framework under development.
\end{abstract}


\section{Introduction}

Testing is an important part of software development, today we count numerous different software testing methods like TDD (Test Driven Development) or BDD (Behavior Driven Development). Those methods are based on the test case writing, allowing automated testing a lot of different situations. 

Behavior-Driven Development (BDD) is an agile software development methodology that facilitates effective communication among individuals with varying levels of technical expertise (developers, quality assurance experts, etc) \cite{north-2006, bellware-no-date}. Indeed, BDD employs natural language to write the test cases, providing detailed descriptions of behaviors and the expected outcomes \cite{north-2006}.

In the world of video games, the process is typically different. Game developers rely on human testers who are provided with a list of elements to evaluate \cite{politowski-2021, bergdahl-2020}. These testers then communicate their findings and results back to the developers. 

Reinforcement learning is a machine learning technique in which an autonomous agent learns to make decisions based on its experience, with the goal to maximize the cumulative reward \cite{kaelbling-1996}. This concept has found applications in various domains, including the gaming industry. For example, OpenAI has developed an agent capable of playing Dota 2 a multiplayer online battle arena (MOBA) \cite{berner-2019}.

Automated testing is already employed by major game developers, employing script-based bots but as more cases arise, the scripted solution's capacity is already being exceeded, necessitating the use of more advanced testing strategies \cite{gillberg-2023}. Presently, research efforts are underway, such as the work of \textit{Gillberg et al.}, who utilized RL with autoplayer bots in their study \cite{gillberg-2023}, and \textit{Gordillo et al.}, who explored playtesting coverage using RL \cite{gordillo-2021}. Nevertheless, numerous video games still contend with a substantial number of issues upon their initial release \cite{politowski-2021}.

\section{Proposal}

Considering the challenges in ensuring the quality of video games upon their initial launch, our proposal outlines an automated testing approach for video games combining the BDD methodology and RL. With this proposal, we establish the following hypotheses:

\begin{hypothesis}{Hypothesis 1}
    The methodology based on BDD for video game testing allows for faster and higher-quality testing.
\end{hypothesis}

\begin{hypothesis}{Hypothesis 2}
    The methodology has the capability to automate various types of tests that are traditionally conducted by humans.
\end{hypothesis}

\begin{hypothesis}{Hypothesis 3}
    The methodology enhances communication between game developers and designers.
\end{hypothesis}

\begin{hypothesis}{Goal}
    Improve the quality of the games through a modification of the game testing methodology.
\end{hypothesis}

To accomplish this, we intend to create a framework and assess it using a game engine, various games, and different reinforcement learning methods as PPO \cite{schulman-2017} or DQN \cite{mnih-2013}.

\begin{figure}[ht]
    \centering
    \includegraphics[width=0.8\linewidth]{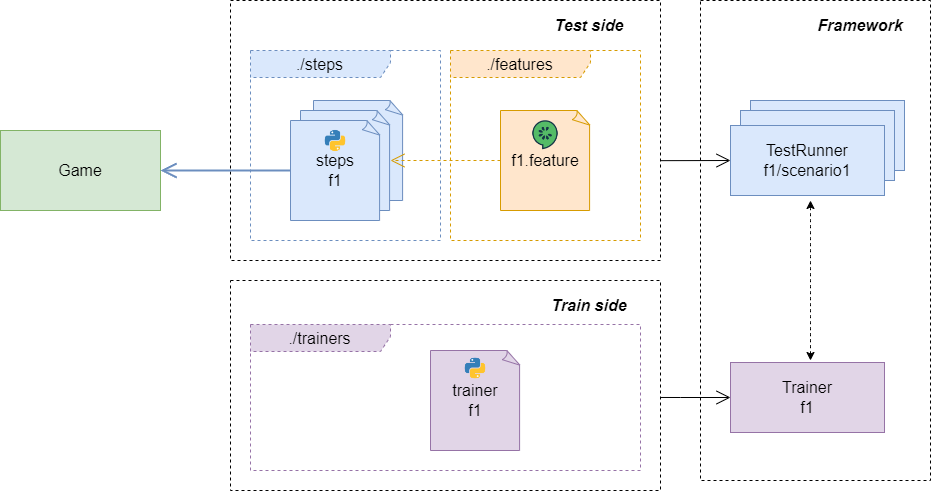}
    \caption{Framework architecture}
    \label{fig:architecture}
\end{figure}

The framework consists of two main components: the “Test side” and the “Train side”. These components are implemented as separate modules within the framework. The figure \ref{fig:architecture} illustrates the architecture of the framework.

By using this method, game designers will write the feature files in the Gherkin language, in our context, a scenario refers to a game situation where we perform the test. Therefore, the "when" keyword of the scenarios denotes a specific moment or a potential situation within the game as opposed to its typical usage where "when" symbolize an event performed. 

\begin{figure}[ht]
    \centering
    \begin{minipage}{0.7\linewidth}
        \centering
        \includegraphics[width=\linewidth]{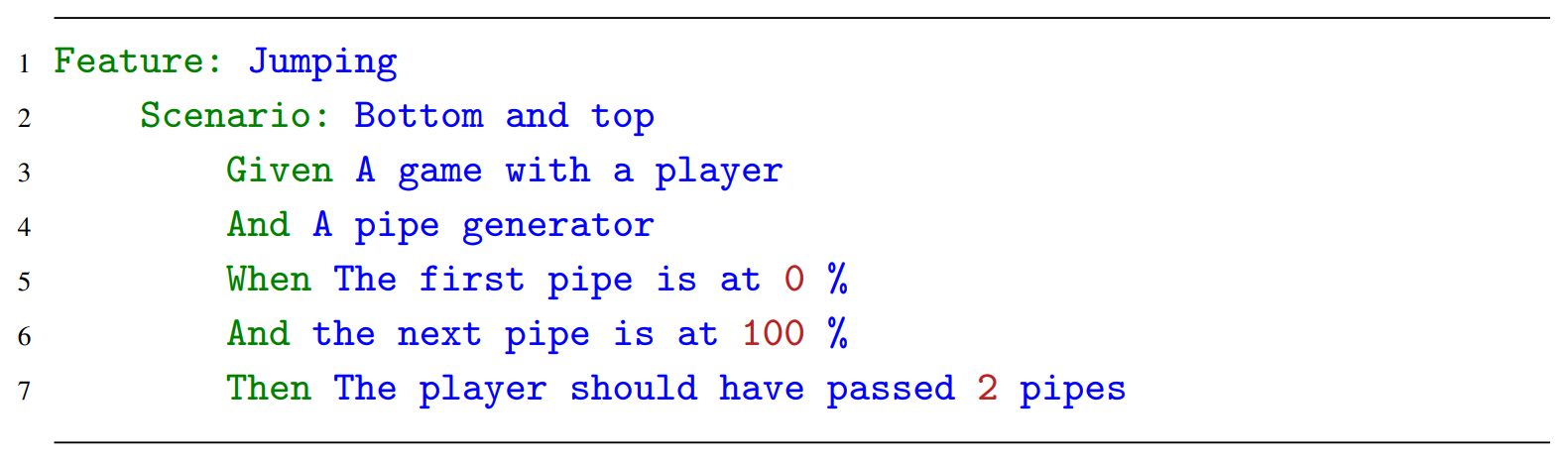}
        \caption{Flappy bird test scenario}
        \label{fig:flappy_bird_scenario}
    \end{minipage}
    \begin{minipage}{0.25\linewidth}
       \centering
        \includegraphics[width=\linewidth]{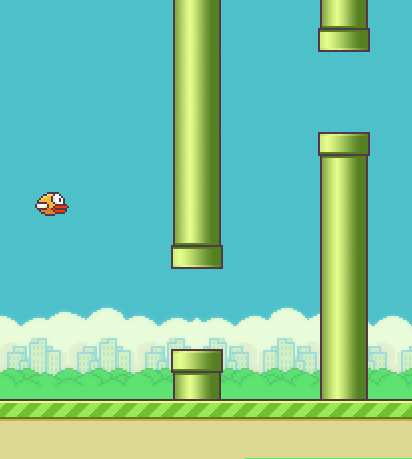}
        \caption{Flappy bird test illustration}
        \label{fig:flappy_bird_test_illustration}
    \end{minipage}
\end{figure}

The following code \ref{fig:flappy_bird_scenario} and the figure \ref{fig:flappy_bird_test_illustration} illustrate the usage of the Gherkin language adapted for video games testing.  In this particular instance, we have employed the popular game "Flappy Bird" for demonstration. The objective of the game is to guide a bird safely through openings between pipes without allowing it to make contact with them. Therefore, an illustrative set of test scenarios would involve verifying the feasibility of the bird passing through various combinations of pipes. In this specific scenario, we are testing a limit case where the first pipe is at the lowest position and the second pipe is at the highest position.

As the usual BDD method the test developer will write the steps files corresponding to each scenario intended by the game designers. They will also write the trainer files which corresponds to the reinforcement learning implementation with a reward function and a feature extraction of the game's observation. 

The framework has two different mode, the train mode which train a model based on the reward function and the game scenario and the test mode which use a trained model in order to pass assertions.

\section{Conclusion}

In this paper, we present our approach of Behavior Driver Game Development combined with Reinforcement Learning and our hypothesis.

In summary, combining the widely used BDD methodology with the growing capabilities of Reinforcement Learning holds great promise for improving the video game development process. This approach has the potential to reduce the workload of video game testers significantly. By leveraging the latest research in Reinforcement Learning \cite{bergdahl-2020, gillberg-2023, gordillo-2021}, we aim to enhance game quality and minimize technical issues upon release, benefiting both developers and players.

\bibliographystyle{unsrt}  
\bibliography{references}

\end{document}